\documentstyle[prd,aps]{revtex}
\begin{document}  
\draft
\twocolumn[\hsize\textwidth\columnwidth\hsize\csname
@twocolumnfalse\endcsname
\title{Truly Minimal Unification : Asymptotically Strong Panacea ?}
\author{Charanjit S. Aulakh$^{(1),(2)}$}

\address{$^{(1)}$ {\it Dept. of Physics, Panjab University, 
Chandigarh, India\footnotemark }}

\address{$^{(2)}${\it International Center for Theoretical Physics,
 Trieste, Italy  }}
\maketitle
\begin{abstract}

We propose  Susy GUTs have a  UV {\it{attractor}}
at $E\sim \Lambda_{cU} \sim 10^{17} GeV $ where gauge symmetries ``confine''
forming  singlet condensates  at scales 
 $E\sim\Lambda_{cU}$. The length   $l_U\sim \Lambda_{cU}^{-1}$
characterizies the {\it{size}} of gauge non- singlet particles yielding a picture
dual to the Dual Standard model of Vachaspati. This  Asymptotic Slavery (AS)
fixed point is driven by  realistic  Fermion Mass(FM)  Higgs content which 
 implies AS. This  defines a  dynamical morphogenetic
scenario dependent on the dynamics of 
UV strong N=1 Susy Gauge-Chiral(SGC) theories.
 Such systems are already understood  in the AF 
case but ignored in the AS  case. Analogy to the 
 AFSGC suggests the perturbative SM gauge group of the Grand Desert confines at
GUT scales i.e GUT symmetry is ``non-restored''.  Restoration before 
confinement and self-inconsistency are the two other (less likely) logical
possibilities. 
Truly Minimal (TM) $SU(5)$ and $SO(10)$ models  with matter and FM Higgs only are
defined;  AM (adjoint
multiplet type) Higgs may be introduced for a Classical Phase Transition (CPT)
description. Renormalizability and  R-Parity leave only the low energy (SM)
data as free parameters in the TM (Quantum PT) case.
 Besides  {\it{ab initio}} resolution of the Heirarchy problem and  
 choice of Susy vacuum, fresh perspectives on
particle elementarity and duality, doublet triplet splitting, proton decay
suppression, soft Susy masses etc open up. 
 ``Elastic'' (spin 2 and spin $3/2$) fluctuations of the  AS  (or {\it{pleromal}})
condensate coupling universally to 
 SM  particles with length scale $l_U\sim l_{Pl}$ imply  an effective $N=1$ 
(super)gravity in the Grand Desert, in which gaugino condensates yield soft
Susy breaking.
A  study of the dynamics of 
ASSGC dynamics to either sustain firmly, or finally dispense with, the dogma 
of  Asymptotic Freedom, is thus required. 

\end{abstract}\vskip1pc]

{\bf A. Introduction.}\hspace{0.5cm} 
{\vskip .2 true cm}

For 30 years Nature has tantalized us with the prospect of
a Grand Unification of all the fundamental interactions of particle physics 
\cite{ps,gg}.
However, inspite of notable steps\cite{gqw75,dimowil,susyrun,einjn1,marsen,amlflo,abmrso10,pross,ewrsb}
 towards the vision of the seminal  works  a truly
convincing minimal and predictive model that reveals the inner 
logic of the  Standard Model (SM) palimpsest has not been singled
out.  The demonstration that the now accurately measured gauge 
couplings unify convincingly only in Supersymmetric models  and
 the discovery of neutrino mass effects
\cite{superk,SNO} (with associated scales near the Unification scale)  
have meanwhile provided a welcome  resilience and elasticity to the
initial vision of symmetry restoration 
at high energy and given us some confidence in the necessity of Supersymmetry.
 They have also driven home the lesson that the SM,
 indeed any good theory, should always
be regarded as an effective theory {\it{a la}} Landau 
useful for coding the regularities of nature apparent at 
some given range of scales of resolution.

\footnotetext{$^*$Email: aulakh@ictp.trieste.it (Till 15 Oct
 2002); aulakh@pu.ac.in (After 15 Oct. 2002).}

Supersymmetric gauge theories based on the gauge group SU(5) and
even more so on SO(10)  \cite{abmrso10} carry our hopes for minimal
and relatively straight forward unification .  On the other hand the
apparent gibberish of the fermion mass spectrum and  vast parameter
 space of the MSSM coupled with stubborn  problems such as Doublet-Triplet 
splitting,  Supersymmetry breaking and suppression of flavour changing 
currents  sometimes make the unification
project seem like a game played by whimsical rules of the practitioners. Meanwhile 
the increasingly stringent constraints on the proton lifetime \cite{superk} have 
engendered both doubts concerning the viability of so
called minimal unifcation \cite{murapierce}   and counter arguments \cite{gorborpav} based on the 
apparent many fold freedom to adjust the parameters of the
MSSM.  Thus only reinforcing current  queasiness regarding the 
whole project . Any additional natural criterion of  minimality of the
Higgs representation choice in GUT models which is well motivated by elementary
  intuitions of dynamical consistency, 
would be welcome. On the other hand a robust and generic indication from within  realistic
GUT models would be equally welcome.
We argue in this letter that such  criteria are  in fact 
visible in the  structure of the SM and implied by it in GUTs.
This Letter is devoted to sketching the picture that emerges intutively leaving
calculational details for later publication\cite{wip}.

{\vskip .2 true cm}

{\bf{B. Massless Charged Fermions}}

{\vskip .2 true cm}
As regards the intuition of dynamical consistency we observe  that : 
{\it{there is no massless fermion interacting with a
 massless gauge field in Nature}} (unconfined ? see below). This   remarkable feature is built into the
  SM by means of a small miracle of
economy. Namely that the {\it{same}}  Higgs multiplet that 
breaks  the $G_{123}$ symmetry down to $SU(3)_c\times U(1)_{em}$  is capable of
ensuring mass to all fermions except the neutrinos (which
are (tellingly !) neutral with respect to both these unbroken symmetries).
Even the discovery of neutrino mass is not incoherent with this economy since 
the SM Higgs can give masses also to neutrinos via  dimension 5 operators \cite{wein,gorakh}.
 This economy of design has perhaps misled us to regard the question of fermion masses as somehow 
peripheral to that of SSB rather than vice versa. 

Usually this economy is attributed merely to the parsimony of Nature 
in choosing the Higgs she needs .  The crucial question on our view ,
 however, is :{\it{ What does She really need to do ?}}  (to build the world we live in).
  Surely it is not {\it{necessary}} to avoid  massless gauge fields as such since these
are the  basis of structure in  the universe. 
On the other hand  Massless Charged Fermions (MCFs)  are
paradoxical objects that can become steadily more energetic in an electric field 
without experiencing any acceleration and {\it{cosi via}}. 
Similarly  QED cross sections (e.g for Compton scattering , pair annihilation etc)
are afflicted with  poles in the electron mass  
showing that the problem  is not merely classical.
Thus  both Classical and Quantum mechanics face notable difficulties in formulating 
well defined physical  theories of such objects. Similar difficulties would 
presumably be faced by massless unconfined YM
theories coupled to strictly massless fermions.  While MCFs  may be
common in some alternative universe with magical 
properties they would certainly be out of place in the quotidian and stable one in
which we live : hemmed in by Iron Laws in Time and Space ! 

Speculations regarding massless quarks may remain consistent with our observation
since $SU(3)_c$ confines in the infrared. 
Confining AF gauge fields may coexist  with MCFs  because  in that
case there are  non-perturbative limits to their spatio-temporal 
freedom and they acquire effective masses in their bag within the  confining Quantum vacuum. In fact
quarks are not appropriate degrees of freedom  for observers ``outside their bag''. 
The example of QCD has further important lessons.  It is now fairly well established 
\cite{vikram} that the Chiral  Symmetry Breaking (CSB) scale 
$\Lambda_{CSB}$ at which the pions are generated is {\it{above}} the  
confining scale $\Lambda_c$. Indeed many valuable insights into Hadron
 Dynamics and structure
have been derived using an effective Lagrangian which couples QCD to 
a linear sigma model of condensate fields (pions ${\vec \pi}$ and $\sigma$).
 Thus there are three regimes in QCD :
I) The confined theory with hadrons and pions . II) The intermediate scale theory
with a quark-pion sigma model just above the confining transition.
  III)  UV theory with weakly coupled QCD and
quarks. The coupling falls steady from IR to UV regimes. It is clear that at high
energies E the IR condensates ($\sim \Lambda_c <<E $) have negligible effects. This is easily
visualized since these energies correspond to violent fluctuations on length scales much
smaller than the IR condensate ones. Now let us turn this picture around and consider a UV
strong QCD -Quark theory. If once again one has three regimes in the same sequence as one
approaches UV confinement the physical interpretation is drastically revised ! In particular
the UV scale condensates being constant on scales much smaller than the wave length at
low energies (much below the UV confinement scale $\Lambda_{cU}$)  {\it{cannot}} be
irrelevant to the dynamics in the perturbative regime  (which in this case is in the IR). In fact the UV condensate 
will define the low energy theory by determining the massless degrees of freedom. The peculiarity of UV condensation however 
lies in that {\it{the labels that are confined at ever smaller scales are precisely the ones that are visible}}
 (to us with our IR eyes).  
We  expect the perturbative degrees of freedom 
(with additions in some intermediate regime) 
to be the correct ones in the Intermediate regime  
(till the usual IR confinement takes over).
Note the peculiar but natural   picture of a quark : a quark would be slathered in glue  
so that it had an effective size $\sim \Lambda_{cU}^{-1}$, but
 on larger scales it's electric flux would stream
out radially since the coupling would be weak :
 a very compact core of glue string tangle horripilating 
field lines like the usual point particle picture at larger scales ! Thus the
role of the AS (or {\it{pleromal}}) condensate is to screen length scales
smaller than $l_U$ from view at larger scales and to thus give a size to the pointlike gauge non-singlet elementary particles 
consisting of a pleromal gauge singlet screen around their putative pointlike center. In other
words the analog of the hadronic bag is beyond
this {\it{pleromal screen}}. This intuitive picture of AS ``confinement'' lies behind our work like a guiding
thread and shows that AS is not ``unthinkable''.

The extension of the above intuitive picture   to the entire SM
embedded in an AS GUT  yields a picture of SM particles intriguingly dual to
the Dual SM (DSM)  of Vachaspati\cite{vachaspati} where the
usual SM particles emerge as monopoles of Dual gauge group with spontaneously
broken GUT scale cores whereas in the present case the cores (also of size
($l_U$)) are instead formed from AS condensates of the usual gauge group. This
fits in well with the usual understanding of Electric Magnetic Duality. Thus the AS in the Electric sector 
implies AF in the magnetic or dual  sector and  justifies construction of the classical monopole solutions basic to the DSM
conjecture \cite{vachaspati}. AF Electric theories would imply large magnetic couplings and thus make the DSM picture less
appealing. Conversely the IR freedom of the Electric theory (at scales $>>\lambda_{QCD}$) implies that 
the Dual theory is IR strong. This may be the reason for the non observation of monopoles.  The topological origin of Electric
charge in the DSM naturally prompts speculation on the nature of the matter multiplets in the SM : could their ``Electric'' charges
also be long distance labels of the strongly coupled electric tangle at their core in our proposed AS picture ? Like the DSM we are now 
on shaky foundations , there is no clear answer to why the mass of such monopoles/tangles is not of the same magnitude as their
inverse size nor how the proper statistics emerges. Perhaps a supersymmetrization of the theory will help in resolving these questions.

As regards the MCF problem one remarks that since in practice QCD is a part of the SM the
answer depends on the running of the EW symmetry breaking vevs which 
give rise to the bare quark masses.  The two alternatives are that  of the well loved 
EW Radiative Symmetry Breaking (EWRSB) \cite{ewrsb}
in which case the EW breaking Higgs vevs will vanish at  sufficiently high scales when the
mass squared parameters of the Higgs doublets turn positive {\it{or}} they do not i.e the EW
vevs are ``hard wired'' into the high energy behaviour. In the latter  case there is 
presumably no MCF problem  in UV free theories . However in the EWRSB case 
one again would have massless unconfined gauge fields coupled to massless quarks at the scale where the 
mass squared parameter turns sign and this we have 
argued is nonsensical/other worldly behaviour. Even if strong coupling screened the problem at the highest
scales the problem  reappears at lower scales unless  at least a seed mass for  
fermions emerges at GUT scales. This again suggests the intimate connection
between 
fermion masses, EW symmetry
 breaking and AS and disfavours purely radiative generation of even the first
generation fermion masses.

{\vskip .2 true cm}
{\bf{C. Truly Minimal Unification}}
{\vskip .2 true cm}

 {\it{Moreover }}  a Higgs sector that is sufficiently complex 
to account  (e.g) \cite{georjarls,babumoha,aulmoh,abmrso10,gorborvis,odatak,oshimo}
 for the observed pattern of fermion masses and
mixings (e.g {\bf{10,120}},${\overline{\bf 126}}$
 in SO(10) GUTs or e.g $ {\bf{(5,{\bar 5})}}$ together with 
  $ {\bf{(45,{\overline {45}})}}$ in SU(5)) implies that as soon as
the the FM residuals'
mass thresholds are crossed (generically this happens
 around  or somewhat above $M_U$) the GUT gauge
coupling grows very strongly and perturbation theory must be abandoned.
 The expectation is naturally that the theory ``confines'' at 
$\Lambda_{cU}\sim 10^{16}-10^{17} GeV $.
However so far this type of behaviour, so
 radically counterposed to the familiar IR confinement, has been
regarded as inconcievable, inconsistent or at best 
to be rescued by  mysterious Quantum Gravity 
Planck foam etc.  By reversing the logic of our expectations and taking
the {\it{MCF problem  as the driving reason for symmetry
breaking}} and the FM Higgs driven condensation as an 
internal message of the inevitability of 
strong coupling/confinement we are then led
to accept the necessity of thinking through
 the consequences of the intuitively new 
``confining'' behaviour in the UV for our picture of elementarity and symmetry
breaking. Fortunately the fact that  Susy GUTs are the favoured unification candidates 
for deep reasons implies that  this  dynamical symmetry
breaking problem  is may be capable  of resolution by the
techniques\cite{afdinsei} already developed and applied 
to the the case of AF theories.  This taming of
the symmetry breaking problem was previously 
applied by us to the case of realistic models with perturbative high scale 
gauge SSB\cite{lrsusy} (actually those theories are also generically AS) :
encouraged, we demand that in a truly minimal theory :

{\bf{(i)}} The Higgs sector should  generate SM fermion
masses.  These Higgs we shall call  {\bf{FM Higgs}}. 
Their vevs must break EW symmetry in an economical and
minimal way to give the observed fermion mass pattern {\it{at the renormalizable and tree level}},
so that the FM-Matter cubic couplings are determined (we shall eschew models with
duplication of Higgs so that there may be a clear 
assignment of functions based on representation structure).   
Other Higgs such as the adjoint multiplets used in SU(5) and SO(10) GUTs 
\cite{ps,gg} or the symmetric/4 index a.s  tensors used in 
minimal Susy SO(10) GUTs \cite{abmrso10,aulmoh}  we shall 
generically call {\bf{AM Higgs}}.
 FM representations are in general complex while AM representations are real in
common models. One typically finds AM representations in the products of FM ones.
The FM Higgs set chosen should of course be anomaly free and this 
provides a welcome further constraint. 

The alert reader will  object that the FM Higgs cannot possibly be
 used to break  symmetry at the GUT scale since then 
$G_{123}$ symmetry also breaks at the GUT scale. We will answer this objection in
detail below arguing that it
need not hold if the strongly coupled theory at the GUT scale generates condensates 
of {\it{products}} of the FM Higgs in  AM channels 
i.e that {\it{the GUT phase transition is quantum}}, while the individual FM fields have vevs that are
much smaller than the GUT scale.
This is made all the more likely by the enormous Casimir indices of typical 
FM irreps beyond the lowest. Inclusion of AM Higgs allows a Classical
 Phase Transition (mean field)
description of the same FM driven condensation and may be preferred
 in practice since the whole trend of
the argument is that the extractable physics lies in a Universality Class
determined by the FM Higgs
content.

{\bf{ (ii)}} We  expect and accept that  at high energies the FM Higgs   
drives the  theory  into the strong coupling regime  where large gauge singlet condensates breaking 
the symmetry down to $G_{123}$  as well much smaller $G_{123}$ breaking corrections develop
(which tame the MCF problem). If this happens much above the perturbative unification scale then only
may we speak of GUT symmetry restoration. However  since symmetry breaking effects are suppressed by
$M_U/ \Lambda_{cU} $ only it is hard to see how a clear distinction could usefully be made. 
Thus we must face the analysis of the ASSGC system head on.

 All, however, is not lost . Firstly low energy fixed
point structures \cite{pross} of the SM and MSSM RG equations for Yukawa and gauge couplings makes them 
insensitive to the high energy dynamics. Secondly behaviour analogous to that of AFSGC theories
would here mean that the perturbative  regime unbroken symmetries   (to leading order in $\Lambda_{cU}$) 
could continue unbroken into the strongly coupled regime as confined while 
the coset ($G_{GUT}/ G_{123}$) gauge degrees of freedom , although now massive, would not be 
decoupled at the the confinement scale since their masses have the same magnitude. The interpolating parameters
being of course the scale of the vev on the one hand and the scale at which the theory is defined on the
other.  Therefore they would contribute to the RG effects which confine the SM gauge symmetries. The same applies 
to other FM residuals ($G_{123}${\it{ non singlets }} ) that became heavy due to the 
 phase transition in the AM channel. However since the UV strong dynamics is, so 
far,  unknown it is marginally  conceivable  that  there some intermediate regime where the
GUT symmetry is neither confined nor broken .
The existence of the GUT symmetry is not  merely a semantic point since the gauge 
and FM residuals would mediate proton decay . The task will be to
 see if these novel ASSGC systems
can yield the Grand Desert type effective softly broken Susy theories with 
very weakly broken $G_{123}$
{\it{and  non zero fermion masses at all scales below}} $\Lambda_{cU} $.
 Notice that this discussion
provides a fresh non-perturbative perspective on symmetry non-restoration
 and thus also on that  route to the resolution 
of the GUT monopole problem\cite{gormoh,dvamelgor}.

Such a UV strong coupling dynamics is already   analyzable to 
some extent due the the great advances in understanding supersymmetric 
strong coupling dynamics \cite{afdinsei} using holomorphy , factorization and single
instanton or dilute instatnton gas techniques. 
{\it{However}} since here we have  UV rather than IR slavery there may be significant
differences between the strong dynamics of these two distinct classes of Susy
Gauge theories.
{\it{In particular the  use of small scale instantons to saturate and thus calculate 
crucial quantities such as the di-gluino condensate may need to be modified because 
the short distance regime is now  also a strong coupling
 regime }}. Indeed a perusal of the arguments used in calculating these condensates quickly
discovers frequent use of the smallness of the running coupling $g(v^2)$ at the 
scale $v$ of the Chiral condensates that define the vacuum. 
 In the present case that regime would be achieved in the region of moduli space where the 
the chiral condensates were {\it{small}} compared to $\Lambda_{cU}$ i.e the intermediate or Grand Desert
region.  The  condensation effects  can still be studied using the general techniques developed but we will
not prejudice the argument by quoting preliminary results \cite{wip} here.
 While this means that no immediate decision on our proposal can be
taken, our arguments will have  served our purpose  if they
motivate   clarification of the nature of ASSGC theories. 
Even if is  rigorously proven that ASSGC dynamics are so 
inconsistent that no such theory should ever be entertained then  so robust 
a support to this tottering Shibboleth of Particle Physics will still be a welcome constraint on 
future speculations.  
 
{\vskip .2 true cm}
{\bf{D.  SU(5) and SO(10) Models}}
{\vskip .2 true cm}

To proceed, let us calculate the possible FM Higgs sets in the case of SU(5) and SO(10) Grand
Unification.  In SO(10) GUTs the fermions of one SM generation (plus $\nu^c_L$)
fit  in a {\bf{16}} plet of SO(10) and mass terms arise from
SO(10) invariant couplings ${\bf{16 \cdot 16\cdot}}${\bf{FMHiggs}} in
 the superpotential. Since ${\bf{16 \times 16 = 10 + 120 + 126}}$  and the
representations{\bf{ 10,120}} are real it immediately follows that the possible FM Higgs in 
SO(10) models are nothing but
 {\bf{10,120,${\overline{\bf 126}}$}}. These have second 
Casimir indices  indices 
1,28,35 respectively. 
R parity preservation requires\cite{abmrso10}  ${\bf{126}}$ to be
paired  with ${\bf{\overline{ 126}}}$.
 The very Higgs ({\bf{120}},${\overline{\bf 126}}$) that are
 used to reach realistic (or quasi)  fermion mass 
relations\cite{georjarls,babumoha,odatak,gorborvis,oshimo} drive the gauge couplings UV strong. 
Since SO(10) has no gauge anomalies any combination of these may  be used.  For the
AM sector (if the reader insists on one!)  one can use, for instance, 
some combination of {\bf{45,54,210}} in SO(10) see e.g \cite{aulmoh,abmrso10}. These also
have a strong effect on the GUT coupling above $M_U$ {\it{ and are all 
contained in the product of the above FM representations}} thus for instance 

\begin{equation} { \bf { 120 \times 120  = 1 + 45 + 54 + 210 +...}}\end{equation}
\begin{equation}  {\bf{126\times{\overline{126}} = 1 + 45  + 210 + ...}}  \end{equation}

In the case of SU(5) we must again
calculate the conjugate of the product of the sum of reducible 
representations ${\bf{1,{\bar 5}, 10}}$ with itself and separate out
anomaly free subsets. This gives : 

\begin{equation}\Phi_1 {\bf{(5,{\bar 5})}}$, 
  $\Phi_2{\bf{(45,{\overline {45}})}}$, $\Phi_3{\bf{(5,{\overline 
{10}})}}$, $\Phi_4{\bf{({\overline{15}},45, {\overline{50}})}}\label{su5FM} \end{equation}

  The SU(5) indices of these combinations are 1,2, 24, 2, 33.  
The 45-plet is the representation used by Georgi and Jarlskog 
\cite{georjarls}. Note that the {\bf{120}} of
SO(10) contains $\Phi_2$ while $\Phi_3,\Phi_4$ lie in ${\bf{\overline{126}}}$
 which can therefore also support the Georgi Jarlskog
mechanism \cite{odatak,gorborvis}. For UV strong coupling 
with just 3 light families one needs at least one of $\Phi_2,\Phi_4$.
In that case (assuming a Grand Desert and sharp FM residuals mass thresholds at $M_U$) 
one finds that  the one loop running GUT coupling
explodes  within  one order of magnitude of the perturbative unifying
scale $M_U \sim 10^{16} GeV$ i.e $\Lambda_{cU}\sim   .7 \times 10^{17} GeV$. 
 For the AM sector one can use, for instance, 
{\bf{24, 75 }} in SU(5) \cite{gg,flipsu5}. They too are UV slave 
drivers and they too are  contained in products of realistic FM Higgs sets.

 We therefore need to, at least, analyze the gauge coupling dynamics of (AM-less) models
like 

\begin{itemize}

\item[I] Susy SO(10) with {\bf{10 + 120}}

\item[II] Susy SO(10) with {\bf{10 + 126}} {\bf{ +}} ${\bf{\overline {126}}}$

\item[III]  Susy SO(10) with {\bf{10 + 120 + 126 + ${\overline{\bf 126}}$}}

\item[IV] Susy SO(10) with {\bf{120 +  126 + ${\overline{\bf 126}}$}}
\item[V] Susy SU(5) with $\Phi_1\oplus\Phi_2$ i.e $\bf{5+{\bar 5} + 45 + {\overline{45}}}$ 

\end{itemize}

and so on : take any anomaly free combination of the FM fields but 
use   {\bf{126}} + ${\overline{\bf 126}}$ in SO(10) FM Higgs sets. See remarks on R-parity
below which justify this . In practice a mean field (CPT) description favours introduction of sufficient
AM Higgs to break the GUT symmetry to $G_{123}$.

{\vskip .2 true cm}
{\bf{E. Actions  }}
{\vskip .2 true cm} 

Following Landau, having isolated  the possible true order parameters of the theory  
the most general supersymmetric action may be specified in terms of an  GUT 
invariant  Kahler potential   $ K(F,{F^*},\Phi, {\Phi^*} , A, {A^*} )$ 
and a super potential $W(F,\Phi, A)$ in a schematic notation in which 
the matter, FM Higgs and AM Higgs chiral superfields are denoted by $F,\Phi,A$
respectively. Naturally these are expansions in powers of invariants of the
 gauge group , with higher powers than 2 in K and 3 in W
being suppressed by powers of a scale $M$ whose exact 
value is left indeterminate for now since we shall focus on renormalizable models.
We have argued that the symmetry breaking and separation out
of the MSSM in the Grand Desert is dynamically determined. The mass scale of FM
residuals(e.g the color triplet Higgs ) is however controllable by the FM mass
parameter thus  opening up a method to resolve the doublet-triplet splitting problem
in a way analogous to models which make the low energy Higgs doublets Goldstone or
pseudo-Goldstone multiplets in some large symmetry scenario.

We remark that we regard R-parity as such a necessary ingredient of a Susy GUT (since
it effectively defines the distinction between matter and Higgs fields, 
and punishes any attempt to mix them with catastrophe) that we
advocate  complete suppression of R-parity violating couplings between 
the F and $\Phi$ and/or $A$ superfields in the case of SU(5) . While for SO(10),
since R-parity preserving GUTs {\it{require}} \cite{abmrso10} a 
${\bf{126 + {\overline{126}}}}$ combination, even though anomaly
cancellation does not ,  we shall use this combination rather
 than a single ${\bf{\overline{126}}}$ i.e the FM fields for
SO(10) will be taken to be ${\bf{10,120,126 \oplus {\overline{126}}}}$. Also  along the same lines
we do not envisage any non-renormalizable terms involving matter fields in the superpotential.
The FM Higgs-Matter Yukawa couplings 
may possibly enter the renormalization of the Kahler potential or into the fine structure (i.e
$O(M_W/M_U)$  and $O(M_S/M_U$))  
of the vacuum state determined  but they will be irrelevant to the
$O(M_U)$ spontaneous symmetry breaking of the
FM Higgs condensates in the AM channels as well as to the  $O(M_W)$  $ G_{123}$ breaking vevs
of the FM Higgs fluctuations themselves and may thus be safely ignored in
a leading order determination of the Gauge-FM Higgs dynamics at $\Lambda_{cU}$.

One  important feature of FM Higgs  superpotentials is that $\Phi^3$cubic couplings
 vanish  so that at the {\it {renormalizable}} level the
superpotential contains only mass
terms e.g ${\bf{10}} ^2 + {\bf{120}}^2$ in model I , 
and  $\bf{5\cdot {\bar 5} + 45 \cdot {\overline{45}}}$ in model V. 
 The extreme simplicity
of the renormalizable R parity preserving pure FM Superpotential ($W= Y FF\Phi + m \Phi{\bar\Phi}$)
 implies that the $Y,m$ are the 
{\it{only}} free parameters besides the values of the low energy gauge couplings,
 gauge boson masses and Newton's constant.
This means that if our scenario is dynamically possible then our Vision
 of the of the {\it{observed}} fermion mass pattern 
 determining the pattern of Symmetry breaking will be realized.

{\vskip .2 true cm}
{\bf{F.  Dynamics}}
{\vskip .2 true cm}

As explained above the current knowledge on Susy gauge dynamics of
 IR strong theories is
{\it{not}} immediately applicable to the UV strong case . Indeed a common reaction 
is to dismiss such a possibility out of hand as too bizzare or counter intuitive.
One argument  is that since a confining theory 
forms a condensate because it seeks to screen the strongly interacting colour 
charges it will not break itself spontaneously in the strong coupling regime.
However as  we have already mentioned AF Susy Gauge-Chiral systems
habitually\cite{afdinsei}
break their own symmetry dynamically  to a subgroup which is perturbative in some 
regions of moduli space and confining in others. This is just the sort of 
behaviour we require and would imply that just the SM
gauge fields confine .
 
Furthermore  the non-lowest FM Higgs representations are 
{\it{enormous}} in terms of  their large second Casimirs.
Thus if e.g fundamentals (such as MCFs!) are confined just above $M_U$ 
(as we argued they should), then even non-singlet combinations of FM fields with 2 or more ``hanging ''
indices 
can be confined all the more since they can still contain many more contracted 
indices than a meson . {\it{Note that once higher FM  representations
  are introduced  quartic and higher 
chiral FM  invariants become possible and are coordinates for the D-flat moduli space. 
Since quadratic products of FM Higgs contain AM Higgs it is certainly 
concievable that a  quartic or higher (even) order  modulus takes its non zero value due to a quantum 
condensation of FM  products in AM channels.}}In order that the symmetry structure and gauge boson mass spectra be 
commensurate the quantum condensates in the chiral ($\Phi\Phi$) and non-Chiral($\Phi\Phi^*$) sectors would need to be 
correlated. 

Another objection is that such a theory would then have no
regime in which its SO(10) provenance could be determined. We cannot see , however, why the
confining phenomena at the GUT scale preclude , for example , proton decay via the massive
bosons with masses $O(M_U)$.
 The above arguments show that  an a priori rejection
 of the possibility of a quantum phase
transition is unwarranted . In connection with the  possibilities of UV strong
dynamics\cite{main} considered in the literature we refer
 the reader to\cite{Rubak} where also   a
preonic    model of 
strong unfication in which  gauge  couplings first 
become strong and then weak again at still
higher energies was described. Presumably the above behaviour was there 
deemed attractive for reason
of the  prejudice that the asymptotic dynamics must be free . 

Our models await rigorous analysis of UV strong gauge theories and their very novel 
{\it{but not necessarily counterintuitive }} features\cite{wip}. 
To illustrate, very schematically, the possibilities we  use the well known
\cite{afdinsei} AFSGC SU(5) model with one
$\phi({\bf 5})+ {\tilde\phi}({\bf{{\bar 5}}})$  
pair(in fact an fermion mass giving  Higgs set for the
minimal SUSY GUT but
not enough for UV slavery : this could be remedied
  (without complicating  FM Higgs sector)
by e.g.  increasing the matter content to more than 7 matter families). Then we could  drop
the matter from consideration on R-parity grounds.  However that still leaves us with the
inapplicability of IR strong results. So we emphasize that we are merely using the known
results for the IR strong model to illustrate interpretational possibilities if at least the weak
assumption that the  connection between the strong and weak coupling regime gauge
symmetries is respected. 
This toy  model  is known to have a non-singular 
Classical moduli space consisting of a single
 chiral invariant $A= {\tilde\phi} \phi$ at 
any non zero value of which the gauge symmetry breaks to $SU(4)$  while 9 chiral 
supermultiplets combine with the 9 $SU(5)/SU(4)$ gauge multiplets via the super 
Higgs effect.  Inclusion of quantum corrections can be shown
to correct the the superpotential only to 

\begin{equation} 
W= m {\tilde\phi} \phi + b(g^2) { {\hat \Lambda_{cU}}\over
 ({\tilde\phi} \phi)^{1\over 4} }
\end{equation}

  Note that in the AS case $m \sim \Lambda_{cU} $ implies that the GUT phase transition 
is in the strong  coupling  regime(so that  Quantum efffects
are appreciable). The second term is generated non-perturbatively for symmetry reasons with  only its
coefficient obtained from a constrained 
instanton calculation\cite{afdinsei} (now dubious in the UV)
 in the weak coupling regime   and  
$ {\hat \Lambda_{cU}}\equiv \Lambda_{cU}^{7/2}$.
  In the $m=0$ case the minimum is at $A=\infty$ and when $m\neq 0$ it comes in to
finite values . For  $m=0 $ there is  global R symmetry which breaks spontaneously as
 $U(1)_R\times U(1)_V \longrightarrow U(1)_V$
and so leaves a massless Goldstone chiral supermultiplet which we can liken to a proto-low 
energy sector. Since the R symmetry is violated at $m\neq 0$ the corresponding 
Goldstone supermultiplet  becomes massive with mass $\sim m$ .
Then   $m\neq 0$  makes the  Goldstone massive since the mass term violates the R
symmetry (softly). However in models with more than one pair of fundamental and
anti-fundamental there is a global $SU(N_f) \times
{\tilde{SU(N_f)}}\longrightarrow
SU(N_f)_{vect}$ symmetry  breaking whose $N_f^2-1$  Goldstone chiral
multiplets  do not suffer from this problem and our remarks can be trivially 
generalized to that case. Of course the analogy is still imperfect beacuse these Goldstones
are still singlets w.r.t the unbroken $SU(N_c-N_f)$ gauge group but there is no reason the
true FM multiplet models may not jump this small group theoretical hurdle easily in view of
their multi-index representations. Indeed this would happen quite naturally in view of our
remarks on higher order chiral invariants present in realistic models and the AM channel 
condensation. Details will be given elsewhere \cite{wip}.

{\vskip .2 true cm}
{\bf{G: Gravity and Susy Breaking}}
{\vskip .2 true cm} 

 The existence of a fundamental length $l_U\sim \Lambda_{cU}^{-1}$ associated with the {\it{gauge
singlet}} condensate that precipitates out leaving our $G_{123}$ labelled world as its low energy limit
provides a normal and minimal route to realizing the old
dream of induced or effective (super)gravity \cite{adler}.
This approach-though initially promising- was largely abandoned in the mid
eighties due to difficulties in obtaining the correct sign and magnitude of the induced gravitational coupling and cosmological constant
\cite{khuri} in AF theories. Even worse  it was pointed out that any theory with quadratic and quartic divergences would suffer from
non-perturbative and irremovable ambiguities in the formulae for these basic quantities\cite{david}.
 The same author also suggested that these difficulties
would {\it{not}} be present in a supersymmetric theory. Strangely this
suggestion for curing the difficulty  
does not seem to have been taken up, possibly  from despair at the difficulties of AF theories. 

Our Susy GUTs  are however quite different animals from those that 
were scrutinized earlier. They are both AS and supersymmetric and as such could evade  earlier difficulties. Morever 
the {\it{global}} supersymmetry of the underlying theory will presumably ensure unambiguous expressions for gravitational parameters 
{\it{and}} ensure  the induced cosmological constant remains
zero ! This would be a severe difficulty in a non supersymmetric fundamental theory. If one accepts the natural 
length scale $l_U$ for the size  of the massless composite  graviton 
(and gravitino) that arise in the effective N=1 Supergravity at low energies (given the
basic induced gravity  premise of general coordinate invariance of the global theory in the gravity background ) then 
it is clear that probes at low energies will not be able to resolve their structure.     
In the induced gravity picture \cite{sakharov,adler} the graviton multiplet may be identified
 with the massless ``elastic'' fluctuations (spin 2,3/2) of the
(very stiff) GUT scale (super)condensate and its universal coupling follows trivially from the  singlet
nature of the condensate. Indeed the behaviour in this regard 
has a curious close relation to the QCD based \cite{neemsij} ``strong gravity'' \cite{kpsinha} expected as a result of QCD
condensates within the Hadronic bag. In that picture di-gluon gauge singlet bound states were assumed to furnish spin-2
hadronic ``pseudo-gravity'' interactions  among strongly interacting hadrons. However since gravity is UV strong while QCD is IR strong
the regimes of that proposal do not quite match.  In the present case both the graviton and the gauge theory are in the
weakly coupled regime in the IR.  The condensate
which veils length scales shorter than $l_U$ from the gauge non-singlet perturbative
 world provides the (super)gravity multiplet.
It is indeed satisfying that the known Planck scale and $\l_U$ are so close. 
The long renormalization down to the scales at which $G_N$ has been measured
 could easily account for the small discrepancy. On decoupling grounds the low
energy effective theory would then be $N=1$ Supergravity coupled to a $G_{123}$ perturbative AFSGC system  
\cite{ferrcrgr}. Since the gauge mutiplet condensates (gaugino condensate) couple to an F term in the
Supergravity they can introduce soft Supersymmetry breaking into the low energy effective theory which is
phenomenologically necessary and may also be needed to generate the observed EW vevs \cite{wip}.

{\vskip .2 true cm}
{\bf{H: Concluding  Remarks }}
{\vskip .2 true cm}

The frankly programmatic rather than technical  content  of this letter
 was originally motivated by the  need to 
face the problem of exploding GUT coupling above the perturbative unification
scale in the SO(10) model constructed by us elsewhere \cite{abmrso10} . It is clearly
completely dependent on the determinable dynamics of the precisely defined set of truly minimal
UV strong  SUSY GUTs  we have set out.   That  analysis can draw upon the impressive
machinery for the analysis of supersymmetric vacua developed over the last 2 decades
\cite{afdinsei} when that has been extended to the UV strong case  for
the fairly complicated realistic cases we have suggested .  We hope to have convinced the reader
that our novel reading of the purpose of symmetry breaking in the SM puzzle and the generic and genuine
difficulty of otherwise perfectly sensible Susy GUTs does
warrant a pursuit of these difficult dynamical questions.
Such an analysis alone can confirm or dismiss the  scenario painted above. 

Having accepted the need to think through the consequences of ASSGC dynamics we found that very mild
assumptions of similarity to the known AFSGC dynamics opened up dazzling vistas of
resolving many of the most obdurate problems concerning the  Unification of the Fundamental interactions,
including gravity, in the context of a very pleasing intuitive picture of
elementarity of particles. Many of these had been given up as dead ends in the course of the years, based (it
seems to us) on an untenable assumption that AS dynamics is necessarily meaningless simply beacause it is
AF dynamics that rules the internal dynamics of the hadronic world. Looking boldly into this blind spot
immediately confers the said dazzling visions ! If these Visions are realizable then we will have washed
away much of the motivation for  ``worm wars'' or ``magic carpet'' mathematically inspired speculation
by taking seriously two elementary generic features of Unifying models associated with the SM : which
codes what we know about elementary particles so successfully.

To conclude we have taken  the  obvious  but non-trivial feature  of  the absence
of MCFs  to have a deep significance and to be the true rationale of
symmetry breaking in nature. Accepting the generic representation structure of the FM Higgs we accept
the drastic consequences that ensue and find that in fact they are eminently
sensible intutively and could resolve many
obdurate problems. We thus suggest that the low energy spectrum
and couplings are the ultimate rationale and determinant
  of symmetry breaking at UV scales in Nature. This should satisfy the most
pertinacious of postivistic ``Machian'' objectors to the validity  of the Grand
Unification project.  Specifically , accepting the obvious FM Higgs  
that typically and easily drive the gauge coupling to UV
slavery ( avoiding MCF problem in the UV) 
leads naturally to a scenario of a {\it{Quantum GUT Phase Transition}}. 

{\it{Forse La Natura canta suoi segreti solo sul canale FM }}

{\it{ O forse abbiamo udito solo La Sirena che 'tira noi }}

{\it{ sugli sassi ultra
violetti !}} 

(Perhaps Nature can be heard  singing her Secrets only 
on the FM band. Or perhaps we have only heard a Siren drawing us 
onto rocky  Ultra Violet shoals !).

{\vskip .2 true cm}
{\bf{Acknowledgements :}}
{\vskip .2 true cm} 

 It is a pleasure to acknowledge the warm hospitality of the High Energy Group of the International 
Centre for Theoretical Physics , Trieste where this work was conceived and executed.
 I am grateful to K.S.Narain,V.P.Nair and Hossein Sarmadi 
for very useful discussions and probing questions. I am grateful and indebted
to Goran Senjanovic for introducing me to  and teaching me about the power of decoupling arguments and 
 minimality . I am also grateful to him for his vigorous defense of the sanctity of the achieved AF
understandings. This work would not have been possible without his influence over the years.
 Naturally no claim is made concerning his endorsement 
of what I have done with what I learnt from him. I am also indebted to Vikram Soni for attempting to teach 
me about Asymptotic Freedom and his boot-strapping vision of SM breaking over the years in spite of my
comatose reactions. I thank my 
family : Satbir, Simran and Noorvir for their patience, good cheer and support during the 
{\it{ aestas miraculorum}} in which this work was conceived and executed. 
Finally I am grateful to Francesco Vissani for warm encouragement and hospitality and the Organizers of
the Gran Sasso Workshop on Astroparticle Physics, July ,2002 for an invitation to attend and speak at 
the workshop.

\end{document}